\documentstyle[osa,graphicx,manuscript,fancyhdr]{revtex}
\title{CCD based phase resolved stroboscopic photometry of pulsars}
\author{Jurij Kotar, Simon Vidrih,  Andrej \v{C}ade\v{z} }
\address{Department of Physics,
Faculty of Mathematics and Physics, University of Ljubljana,
Jadranska 19, 1000 Ljubljana, Slovenia}

\begin{document}
\maketitle

\begin{abstract}
A stroboscope designed to observe pulsars in the optical spectrum is presented.
The absolute phase of the stroboscope is synchronized to better than $2.5\,\mu$s
with the known radio ephemerides for a given pulsar. The absolute timing is
provided by the GPS clock. With such a device phase resolved photometry of
pulsars can be performed. We demonstrate the instrument's capabilities with the
results of a set of observations of the Crab pulsar, the brightest of the known
optical pulsars, with a visual magnitude of $16.5$, and a rotational frequency
of $\sim 29\,$Hz. 
\end{abstract}

\section{Introduction} 
In this paper we present an improved version of the stroboscopic system,
previously developed and used for observing faint optical signals from pulsars,
specifically the signal from the Crab pulsar~\cite{Gal_1,Gal_2,Gal_3}.
Pulsars are fast rotating neutron stars~\cite{Smith}. Relativistic processes in
the pulsar magnetosphere give rise to synchrotron emission, which is
subsequently modulated by the neutron star's rotation. This results in a
periodic train of pulses, the latter typically being a small fraction of the
pulsar's rotation period. For pulsars the rotation periods range from several
milliseconds to several seconds.

The main idea of the stroboscopic system is to observe the light signal
periodically only during part of the period. The technique is based on a 
shutter that opens with the prescribed frequency and phase. In our case the
shutter is a rotating wheel with out-cuts whose width determines the fraction of
the period during which a light detector - in this case a CCD camera - is
illuminated. Photometry is usually performed with a chopper with out-cuts
corresponding to the pulse width and with the phase of the chopper synchronized
to the topocentric pulse arrival times of a given pulsar.

The commonly applied method for observing pulsars at optical wavelengths is by
high-speed photometers~\cite{Chakrabarty,Golden,Shearer,Butler}. This choice is
dictated by the required time resolution. Normal CCDs are too slow for the task,
since their typical integration times are several seconds, which may be more
than hundreds of pulsar periods. However, we argue that a precisely timed
stroboscope in front of a CCD camera can make up for the lack of time resolution
and offers the advantage of the higher quantum efficiency\cite{avalanche} as
well as that of a more detailed field of view. It thus enables highly accurate
phase resolved photometry of even fast pulsars. We note, however, that in the
last years the time resolution of CCDs has been increased by the use of frame
transfer technique, so that second \cite{Dhillon} and even subsecond \cite{Kern}
resolution has been achieved. An ingenious method of periodically saving
(linear) spectroscopic data by a CCD has been applied and used by Fordham et
al.\cite{Fordham,Alberto}.

\section{Principle of operation}
From the observer's point of view the frequency of a pulsar is constantly
changing due to the Doppler shift caused by the Earth's relative motion, the
pulsar's spin down, and also due to changes in the local gravitational redshift.
In the case of the Crab pulsar the maximal change of the frequency due to the
Doppler shift is $\sim \pm 5\times10^{-5}\,$Hz, while the first derivative of
the frequency due to the pulsar spin down is $\sim -10^{-14}\,$s$^{-2}$. All
these effects are calculated with very high accuracy. Using the standard program
TEMPO~\cite{Tempo} we calculate topocentric arrival times of pulses using known
radio pulsar ephemerides and geographic coordinates of the observatory. Absolute
timing provided by the GPS clock assures an accurate reference to the absolute
radio phase of the pulsar. 

The electronic system described in the next section drives the chopper so that
it is phase locked with a given pulsar's ephemerides, which are derived from
continual radio timing studies\cite{Jodrell}. Thus one can examine the relation
between the radio and the optical phase of the pulsar, search for other unknown
periodic changes of the pulsar signal on a time scale longer then the
integration time of a CCD~\cite{Vidrih}. One can also change the chopper phase
in prescribed steps and thus investigate the shape of the pulse signal. Examples
of such measurements together with stability tests are presented later in the
paper.

\section{Instrument}
\begin{figure}[h]
\begin{center}
\includegraphics[scale=0.8]{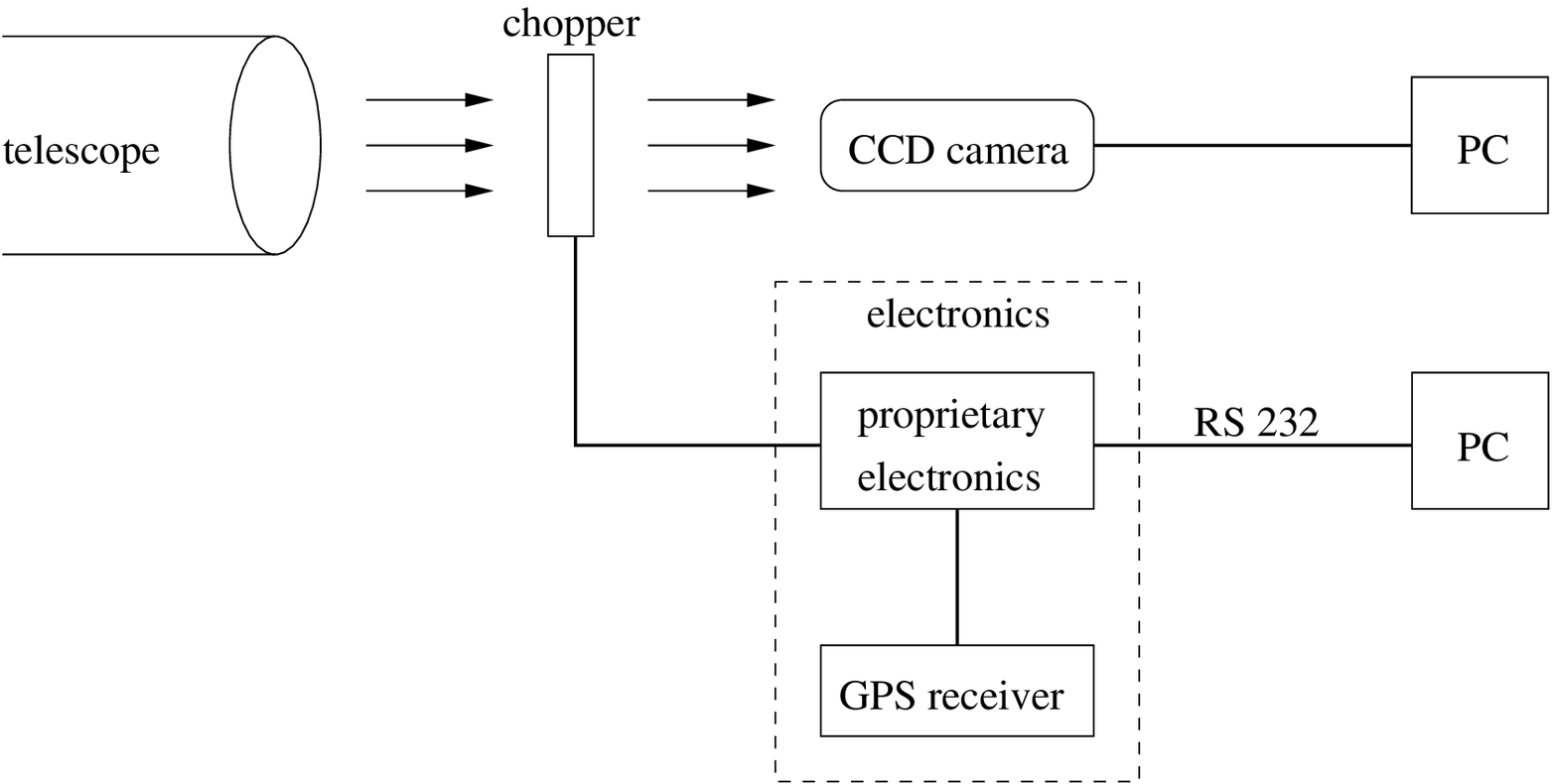}
\vspace{0.5cm}
\caption{\label{schemesystem}The sketch of the stroboscopic system. The CCD
camera is independently connected to another PC.}
\end{center}
\end{figure}

As shown in Figure \ref{schemesystem}, the system consists of a chopper,
electronics for driving and controlling the chopper and a PC, connected via a
serial port to the electronics.

\begin{figure}[h]
\begin{center}
\includegraphics[scale=0.8]{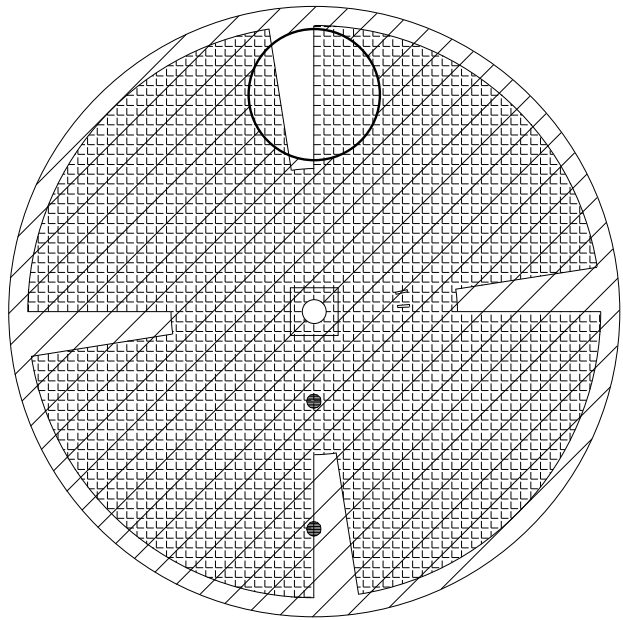}
\vspace{0.5cm}
\caption{\label{schemechopper}The drawing of the chopper. The light passing
through the hole in the enclosure is chopped with the rotating blades. Optical
sensors on the enclosure (marked as two black spots) measure the position of the
blade once per revolution.}
\end{center}
\end{figure}

The chopper is made of a steel blade $240\,\mathrm{mm}$ in diameter which has
four out-cuts (Figure \ref{schemechopper}). The chopper blade is enclosed
in a case and rotated by a stepper motor with 500 steps/revolution. When
observing the Crab pulsar, it spins with one quarter of the Crab pulsar
frequency which is approximately $7.5\,\mathrm{Hz}$. The width of out-cuts is 9
degrees, which corresponds to the width of the Crab pulsar main pulse (see
Figure \ref{crabprofile}). During observations the chopper is mounted in front
of a CCD camera on a telescope. Optical sensors on {the} chopper housing are
used to measure the phase of the blade once per revolution. 

{The} electronics consists of the Truetime GPS receiver, model XL-DC
151-600-110~\cite{Truetime} and a set of proprietary electronics components
(home made DDS board, stepper motor driver and microprocessor board).

{A} GPS receiver generates  a highly accurate clock of $10( 1 \pm
3\times10^{-12})\,\mathrm{MHz}$, which is used as a frequency reference. Two
other functions of the receiver are used {to}: i) generation of short pulses at
specified absolute time (used for frequency and phase control of the chopper
blade) with an accuracy of $150\,\mathrm{ns}$, and ii) the absolute measurement
of arrival times of externally applied input signals which are generated by
optical sensors on the chopper.

{The} DDS board is based on  Analog Devices' AD9854 digital
synthesizer~\cite{Analog} which provides 48-bit frequency resolution. It
generates clock output with resolution of  $3.55\times10^{-8}\,\mathrm{Hz}$ from
$10\,\mathrm{MHz}$ frequency reference. The clock output drives the stepper
motor by means of two National Semiconductor's LMD18245 stepper motor driver
chips~\cite{National}. The microcontroller board based on Atmel AVR8515
microcontroller~\cite{Atmel} is responsible for communication between DDS board,
stepper driver, GPS receiver and the PC.

The system is controlled from the PC. First ephemerides for the pulsar phase are
calculated by TEMPO and loaded in a control program which first sets the chopper
to run with the frequency suitable for the starting time. Next the phase of the
chopper with respect to the pulsar ephemerides is measured and calculated. This
is done by repeatedly reading the GPS clock when the chopper out-cut passes the
beam of light between a LED and a photodiode mounted on the enclosure (see
Figure \ref{schemechopper}). After the absolute phase of the chopper has been
established, the correction to the desired phase is calculated and the chopper
driving is taken over by the computer to follow the phase, according to the
output of the TEMPO program.   

Here we should like to emphasize that our system relies on pulsar ephemerides,
so that all accuracy measurements presented here are with respect to the
ephemerides phase. In practice  radio ephemerides for pulsars is not always very
accurate. For example the Jodrell Bank ephemerides for the Crab pulsar is known
to have an absolute phase uncertainty between 40 and 1000 microseconds. The
relation between the radio and optical phase has not been extensively studied.
However, from our (albeit limited) experience we can state: using Jodrell Bank
ephemerides that is no older than 1/2 month and carefully measuring the phase
with respect to the ephemerides on the first day, we find that during the next
day of observations the peak brightness is obtained at the same relative phase.

\begin{figure}[h]
\begin{center}
\includegraphics[scale=1.3]{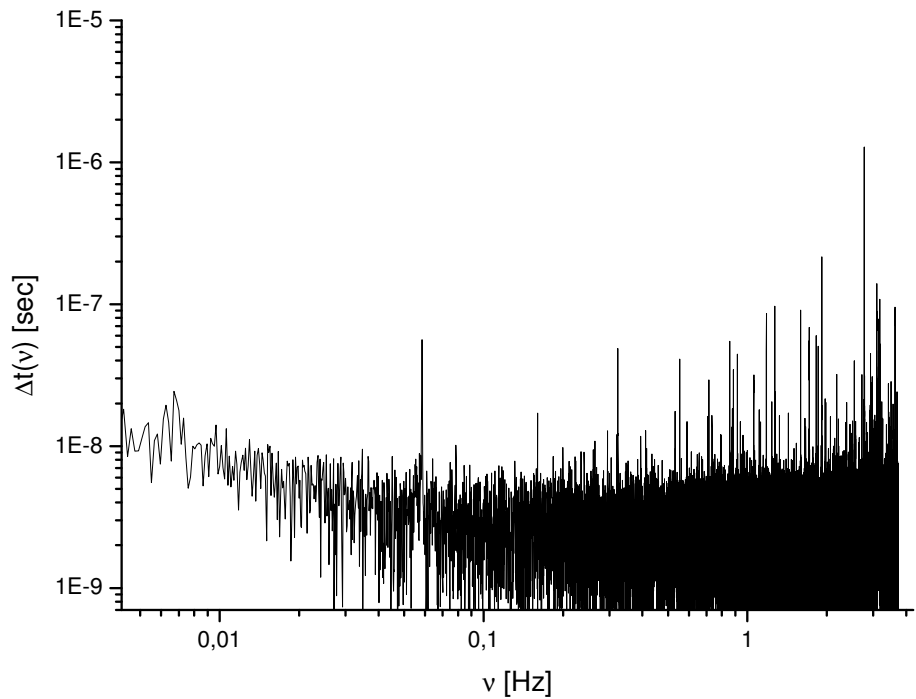}
\vspace{0.5cm}
\caption{\label{freqsp}Frequency spectrum of timing noise. $\delta t(t_i)=\bigl
[\int_0^{t_i^\prime} \nu(t) dt -i\bigr ]/\nu(t_i^\prime)$, where $t_i^\prime$ is
the time of detecting the i-th passage of the chopper through the LED trigger
and $\Delta t(\nu)$ is the Fourier transform ${1\over T}\int_0^T \delta t(x)
e^{-2\pi i \nu x}dx$.} 
\end{center}
\end{figure}

The phase tracking accuracy depends on two circumstances: i) on the accuracy of
the absolute phase $\psi$ of the chopper with respect to the housing and ii) on
the position of the beam from the pulsar with respect to the housing. The
accuracy of the absolute phase is limited by timing noise, which we define as
the difference between the time when the chopper actually opens and the time
when it is expected to open. We measured the spectrum of this noise and found
that it is mainly due to random excitations of mechanical resonances in the
motor-chopper system (see Figure \ref{freqsp}). Its RMS value was found to be
2.5$\,\mu$s. The phase dependence ($\delta \psi$) on the position of the pulsar
(position of the pulsar's image on the CCD) with respect to direction of the
optical axis of the telescope ($\delta \phi$) is easily calculated from the
geometry of the beam and is: 
\begin{equation}
 \delta \psi = {4 F\over r}\delta\phi~~~~,
\label{ }
\end{equation}
where $F$ is the effective focal length of the telescope and $r$ is the radius
of the chopper. For the optical setup in Cananea this turns into $\delta \psi
\approx 0.0054 {rad\over "} \delta \phi$. In order to avoid possible phase
modulation by the periodic error in the telescope tracking, the chopper blade is
mounted so as to cut into the beam in the north-south direction. 

The phase guiding routine works as follows: the position angle of the chopper
($\psi_i$) at setting moments ($t_i$)\cite{SettingMom} is calculated as the sum
of the initial phase and the phase increment, which is the integral of frequency
corrections (DDS is particularly suitable for the task, because it is built in
such a way that the result of integration is exact). The phase in the following
moment ($t_{i+1}$) is calculated from TEMPO and the frequency necessary to reach
this phase is calculated and set. The phase guiding routine is checked at will
by additional measurement of the blade phase with GPS timing as described above.
We thus verified in many hours of running that the phase guiding routine
preserves the average (40 successive measurements) phase of the blade to within
$0.5\,\mu$s for several hours.

\section{Experimental results}
The stroboscopic system has been successfully tested during observations of the
Crab pulsar at the 2.12 m telescope of the Guillermo Haro observatory in
Cananea~\cite{GH} in January 2002. Since the Crab pulsar light curve is well
known from other observations~\cite{Gal_3,Alberto,Percival}, one of the tests
was to measure the light curve again and compare it with those previously
obtained. 

All the images were obtained with the EEV P8603 CCD detector and no filters.
Thus the optical range of observation is approximately between 500 and 900 nm.
During the observations biases and dome flats were taken in order to correct for
the CCD bias and gain imperfections. The fluxes were calibrated with respect to
the photometric standards PG 0942-029 and PG 1047+003 (Landolt\cite{Landolt}).
Note, however, that images were taken without filters, so the calibration is
only relative. 

The light curve was measured during the early night hours of January 9/10 2002.
Sky conditions were changing with some scattered clouds in the sky. We covered
the whole phase light curve with the set of 36 images $1/36 \times 2 \pi$ apart
in phase. The exposure time of each image was 45 seconds. The pulsar was
detected in 18 frames and was too faint in the remaining 18 images. The results,
shown as crosses in Figure \ref{profile} (the width is much larger than the
actual phase uncertainty), agree with other measurements of the optical light
curve\cite{Alberto,Percival}. The solid curve in this Figure shows the optical
light curve obtained by Fordham et al.\cite{Alberto} convolved with our
stroboscopic window function. We fitted our stroboscopic data to this curve
allowing the phase difference and the flux scaling parameter to be determined by
the fit. In this way we obtained\cite{NumericalRecipes}
$\chi^2/(N-M)=\frac{140}{16}$. The $\chi^2$ has been calculated assuming that
the only noise is in our data and has the uncertainty given by
{IRAF/DAOPHOT\cite{IRAF,Daophot}}, which is the uncertainty due to photon shot
noise only. The additional systematic error in determining the flux from the
pulsar which comes from its position in the middle of the nebula and from the
neighboring comparably bright star has not been included in the calculation of
$\chi^2$. By examining the data, we verified that the magnitude of the
neighboring star is anti-correlated with that of the pulsar, making the
variations of its flux about 2.5 times greater than the IRAF/DAOPHOT calculated
value. Thus we conclude that our data agree with Fordham et al.\cite{Alberto}.
Note that the amplitude of the systematic error discussed above is proportional
to the ratio of maximum to minimum pulsar flux in the set of measurements, so
that it becomes negligible for fixed phase photometry.

\section{Acknowledgments}
\begin{figure}[h]
\begin{center}
\includegraphics[scale=1.3]{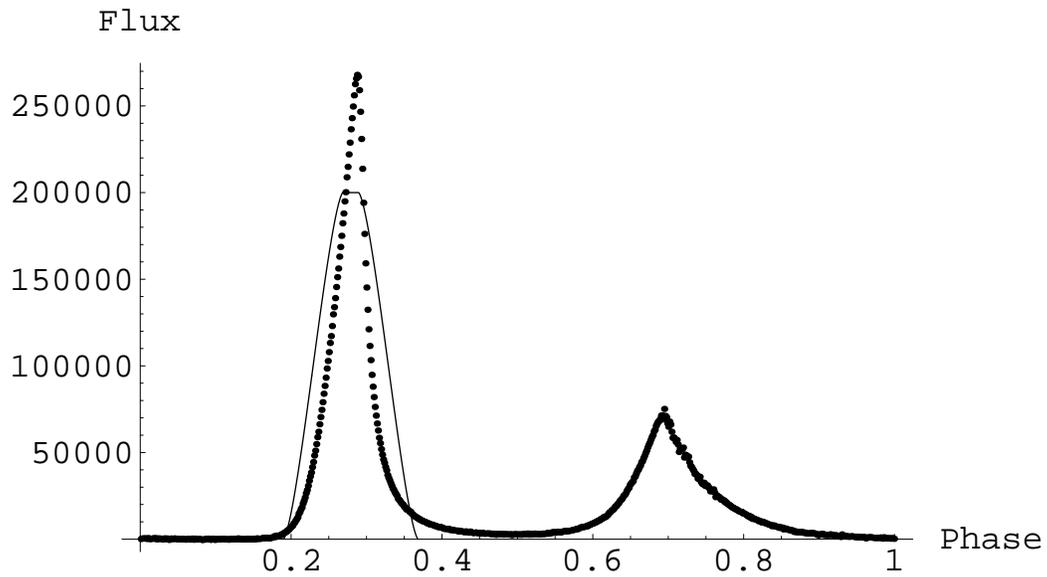}
\vspace{0.5cm}
\caption{\label{crabprofile}The light curve of the Crab pulsar, obtained by
Fordham et. al ~\cite{Alberto}. The width of the main
pulse is $\sim 10\%$ of the whole period. The stroboscopic window function is
also displayed and has been calculated on the basis of the width of the chopper
blade out-cut and the position of the chopper blade with respect to the focal
plane of the telescope.}
\end{center}
\end{figure}

\begin{figure}[h]
\begin{center}
\includegraphics[scale=0.8]{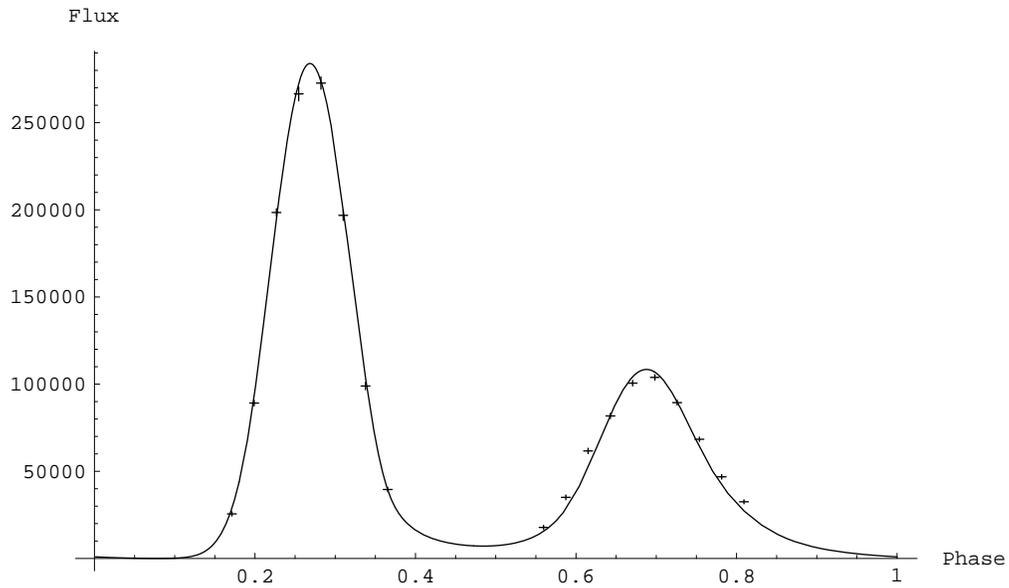}
\vspace{0.5cm}
\caption{\label{profile}Fitting the stroboscopic phase light curve to
\cite{Alberto} phase light curve data. Data from Fordham et al.\cite{Alberto} 
(see Figure \ref{crabprofile}) convolved with the chopper window function are
shown as a smooth curve. Corresponding stroboscopic fluxes together with flux
errorbars, as given by IRAF/DAOPHOT, are shown as crosses. In finding the fit
only the absolute phase and magnitude shift are considered as free parameters. }
\end{center}
\end{figure}

We are grateful to Alberto Carrami\~{n}ana who made observations of the Crab 
pulsar  possible and to Carrami\~{n}ana, Fordham et al. who gave us data on the
Crab pulsar phase light curve (see Figure \ref{crabprofile}) before publication.
We thank the technical staff of the Cananea Observatory for friendly atmosphere
and expert mounting of the instrument. Igor Poberaj gave the DDS board he
developed. We thank Dick Manchester and David Nice for help with TEMPO.
Du\v{s}an Babi\v{c} criticized the text and made it more readable. We would also
like to thank the anonymous referee whose comments and suggestions improved the
presentation of this paper. This research was supported in part by the Ministry
of Education Science and Sport of the Republic of Slovenia.

\newpage

\newpage

\newpage

\begin{thebibliography}{99}
\bibitem{avalanche} The quantum efficiency of newly developed Avalanche
		Photodiodes is comparable with the quantum efficiency of a
		CCD~\cite{Straubmeier}.
\bibitem{SettingMom} Setting moments are started by GPS periodically at 	
			$t_i=i \Delta T$, where $\Delta T$ is usually a prime number of
			seconds.
\bibitem{Daophot} {DAOPHOT} is a software package for doing stellar
	photometry in crowded fields developed by Peter Stetson\cite{Stetson}.
	The IRAF/DAOPHOT package uses the task structure and algorithms of
	DAOPHOT to do crowded field photometry within the IRAF data reduction
	and analysis environment.
\end{thebibliography}
\end{document}